# Hidden human variables in quantum mechanics?

Göte Nyman

Department of Psychology, Faculty of Medicine

00014 University of Helsinki, Finland, P.O. Box 21

(Contact: gote.nyman@helsinki.fi)

The problem of the observer in quantum mechanics is getting new human content. The paradox of Wigner's friend and its extended versions have observers who not only observe quantum phenomena, but communicate, have memories and even super-observer powers. Observers are represented by particle paths and state memories and advanced AI has been suggested to act as an observer. There is a new call for a solid theory of the observer in quantum mechanics. Two different branches of observer theories have emerged. The purely physical one is Heisenbergian, e.g. relational quantum mechanics where the observer is considered as any other physical quantum system. The other branch is psychologically rich as its observer has complex human faculties such as a mind, mental states and memory (Many minds), or the observer is considered as an active and experiencing agent, with continuously refreshed (scientific) expectations (QBism). Is the human observer a special case, different from other quantum mechanical systems? Why is there no theory of the general observer in quantum mechanics? A historical summary is covered on how quantum mechanical interpretations have treated the observer, and the concept of 'hidden human variables' is suggested to continue the discussion on the theoretical nature of the observer.

## OBSERVATION, MEASUREMENT, AND INTERPRETATION

Can we observe a quantum, directly with our own eyes? No, probably never. Of course, it is possible to generate signals from quantum measurements, so that the experimenter can see, hear, or even feel them by using simple recording and transformation tools. When Max Planck (1900) introduced the equation E=h$\nu$ he hypothesized the packets of energy behind it [1] but did not suggest direct methods for observing these 'formal' energy-carrying elements. Einstein (1905) then made the explicit suggestion that quantum behavior can be experimentally observed. It was possible to take one step closer to observing light quanta as localised objects: *"[light ray] … a finite number of energy quanta, localised in space, which move without being divided and which can be absorbed or emitted only as a whole"* [2].

Can we see light waves? Thomas Young [3] observed the fringes, i.e. their spatial frequency on the screen and could vary their appearance by adjusting the physical parameters of the set-up. This allowed the rational *interpretation* that the fringes were produced by interacting waves of light. Nevertheless, we cannot see light waves directly although we see colors which are pure subjective sensations emerging from the activation of the wave-length selective retinal cells by light - or by any other effective energy that can stimulate them. Further, due to color contrast and adaptation, a perceived color does not carry unique or accurate information about the wavelengths of the light source. We cannot see light waves, either. What *can* we see then?

In the first heated debates on the role of the human observer in quantum mechanical measurements, visual perception was typically implied. Theoretically, however, a complete quantum mechanical world model requires *a general theory of the observer*, not limited to specific human faculties. Such a formal theory has not been suggested and implicit assumptions about the human observer hide in the well-known interpretations of quantum mechanics, e.g. Copenhagen [4], Many worlds [5], de Broglie-Bohm [6], QBism [7] [8], in Wigner's quantum paradox [9] [10] and in its extended version [11] [12].

According to the orthodox Copenhagen interpretation, there is the inseparable interaction between the observables and the measurement apparatus; the human observer can only conduct measurements and read indicators. Some may question this impotent role of the experimenter by reminding that a human observer can *detect* the visual impact of a single photon [13] [14] and an isolated receptor of the frog retina reacts to single photons [15]. This has even encouraged to speculations that quantum superposition could exists in the eye [16], but so far, it is not known, what would be a complete quantum theoretical model for describing the state of even the first layers of the visual system, the retina.

**Human, inanimate or theoretical observer?**

The human observer is essentially an intelligent interpretation system, different from inanimate systems, the most advanced AI included. An observer who detects the impact of a weak light or a single photon has only a minimum amount of information for interpretating the sensation: she becomes aware that *something* has caused the sensory experience – it is a *state change* in the observer, but that is all. Hence, a 10 ms duration, low-contrast photograph of a cow and a random flow of photons, for example, can appear identical to the observer, who does her best to interpret the sensory experience-state.

It is somewhat obscure what exactly is meant by the theoretical and especially human observer in quantum mechanics and the problem is getting ever more enchanting. Recent thought experiments include super-observers with strange powers, and experimenters observing each other, communicating and having erasable memories. Advanced AI and photon paths or photonic memories have been suggested to represent the human observer and microscopic and macroscopic (observer) systems are treated as quantum mechanically equal [9] [11] [12] [17]. There is a call for a solid physical theory of human and inanimate observers. At the moment it is not clear at all where and how such a theory could emerge. The general observer has remained totally outside the camp fires of theoretical physicists – and perceptual psychologists.

**What do quantum theorists talk about when they talk about the observer?**

The terms sensation, measurement, perception, observation and interpretation were used haphazardly by Einstein, Bohr, Heisenberg, and Schrödinger. Einstein, in the spirit of the Lorentz invariance, trusted that perception mechanisms remain invariant in extreme conditions, and he used everyday language to describe the subject's relative perceptions when speeding at close to the speed of light. For Bohr the term 'observation' meant more than the act of a single observer in the laboratory and he used the term as a synonym for 'measurement'. He saw everyday language as necessary for expressing the human role in physical observations [18] [19]. For a perceptual psychologist, Heisenberg sounded harsh: *"It does not matter whether the observer is an apparatus or human being…"* [20]. Schrödinger, in his "ganz burleske" quantum-cat metaphor, used the term 'direct observation' (direct Beobachtung), to describe the experimenter's observation of the cat in the box [21]. What exactly this 'direct' meant was left unclear. Bell even suggested that the word 'observable' should be banned from exact formulation of a physical theory [22] and he offered an alternative concept, 'beable': *"The beables must include the settings of switches and knobs on experimental equipment …'Observables' must be made, somehow, out of beables."* [23]

The hypothesis of wave-particle duality was not welcomed among the theoretical physicists of 1900-1920s [24] and the problem of the observer remained open. In philosophy it had been a recurring topic [25] but it took some time for the observer problem to find its quantum-theoretical position: What quantum phenomena exist to be observed? What is and what is not an observation? How to formally connect the human observer with quantum phenomena? The burning problem remains: how do observations and perceptions inform us (humans) about reality?

The early quantum theorists were agnostic to the detailed mechanisms of human perception although psychophysics had already found its scientific roots [26]. Fechner worked with the classic complementarity problem: that of mind and body and saw them as different sides of one reality. Considering the inherent links of psychophysics to physics, it is surprising that no formal psychophysical theory of the observer emerged to challenge the Copenhagen interpretation in its early days.

In 1996 Rovelli introduced the relational quantum mechanics interpretation, which evaded the historical measurement/observer problem in the Heisenbergian spirit by offering the equivalence hypothesis that an observer is like any other system and should not be treated as a special case or as including a human being at all: *"All systems are assumed to be equivalent, there is no observer-observed distinction"*. Any macroscopic system, living or not, could then be considered as an observer, and furthermore, two observers can have different observations of the state of the same quantum system. Reality can only be coded relative to the observer and (quantum) communication is needed between the observers who want to share their different views about the same event. [27].

QBism [7] [8] takes a strong subjective view to quantum mechanical theories and methods as means for the human *agent* to formulate actions and subjective, probabilistic expectations, however scientific they may be. This then leads to updated knowledge and new expectations. Where Rovelli's human observer is no different from other systems or even from a table lamp, QBism includes a participating and experiencing human agent. 'Observation' is transformed into a complex problem of action, observation, interpretation, and collection of personal experiences by the experimenters and sharing them with the members of the scientific community [28]. Quantum formulation of such complex human phenomena is problematic, if not almost impossible.

**Hidden human variables for ever?**

Referring to the EPR paradox [29] Bohm used the concept of "hidden" variables [6], which should be known in order to make 'the usual interpretation' of quantum theory complete. Knowing these additional physical variables would allow prediction of the precise behaviour of a deterministic quantum system. EPR had emphasized the correspondence hypothesis: *"In a complete theory there is an element corresponding to each element in reality."* The human observer belongs, of course, to this physical reality, and the correspondence demand must concern her as well. The problem is not made easier bearing in mind that the definitions of physical 'elements' are products of the human agent having limited sensory-perceptual resources.

We can enjoy the idea of pure physical observers as quantum systems and even a table lamp being in atom-scale interaction with other objects - as Rovelli suggests – but a lamp making interpretations of the world and communicating with other objects calls for a strong quantum theory of human and inanimate communication. One reason for the obscure definitions of human observation in physics is that theories and research paradigms of human behaviour have remained distant to quantum mechanical formalisms. Studies on photon vision and a few explicit treatments of observer mechanics have been directly aimed at quantum mechanics, e.g. [30] and 'quantum cognition' approaches have looked at the quantum phenomena from a higher, rather speculative cognitive-conceptual viewpoint [31].

QBism is perhaps the strongest psychologically grounded theory of physics, but there is no formally complete perception-experience-action theory that could be directly applied to it. Neisser's classic, qualitative model of the perceptual cycle, for example, comes close to the overall framework of QBism. There the observer has a mental model that guides his observation and information search, which in turn leads to updating the *mental* model and expectations [32].

The physical methods to measure length, time and mass hide complex human variables: the measures were originally developed to compensate for the inaccurate observation-performance of humans. The story of the meter, including the human errors and fraud in defining it, is an amazing example of these human powers [33]. Fechner developed the experimental methods and algorithms for quantifying human perceptual abilities and deficiencies. In other words, *hidden, but measurable human variables* were introduced, from the start, into all physical concepts, measures and means of observation. They still hide there [34] [35], only different from what Bell [36], Bohm [37] and the EPR team [29] considered.

## WHAT CAN A THEORY OF THE GENERAL OBSERVER TELL TO THE INTERPRETATIONS OF QUANTUM MECHANICS?

Is the hypothesis of human hidden variables only a wild thought experiment? We know that the interpretations of quantum mechanics [5] [7] [27] [38] are built on human-centric physics and include strong, implicit assumptions about the observer. Everett even made the direct suggestion that the observer has specific subjective faculties: "… *an observer (state) with subjective knowledge (i.e. perceptions)*" [5], but he did not formulate them in any detail. In the many minds model [38], each mind/observer has mental states, experiences, a memory, and beliefs and hence, as observers, they have non-sensory capabilities for which there is no complete formal system description available. Non-sensory, subjective variables have been included in the 'agent' to make her more than a pure observer system [7][18]. These human aspects are familiar from the early theories of active perception, e.g. by Gibson [39] and can be found in interpretative perception concepts, e.g. [40] [41]. Their explicit role in quantum mechanical theories should be defined and formulated.

There is no theory of the general observer in quantum mechanics or in perceptual psychology either. Instead of imagining a dead table lamp [27] or *"any system that can extract information from another system"* [12], a thought experiment including a living frog with frog eyes but a human brain is instructive [34]. Non-locality would be a natural phenomenon for this brainy creature who – because of its derivative and non-linear eyes - does not 'see' a static meter stick in front of its eyes unless the stick or the frog is wildly waved back and forth; the human meaning of 'distance' would not exist for it, or at least it would be computationally very different. Generalizing from this, any number of different observers can be imagined, with their own peculiarities like 'sensory dimensions' we humans don't have, or lacking those that we have. The human observer is a special system with its interpretative capacities and the specific perceptual constraints that originate from her observation mechanisms. Is it possible then, to step out from the scope of the human-centric physics, to imagine and formulate other observers and see the universe with even slightly different eyes or whatever 'observation channels' these might be? Can the notion of a general observer lead to any tangible experimental predictions? Does it make theoretical sense? Whatever the answer to this bizarre, new-age sounding enigma, complex human variables have entered the room of quantum mechanical interpretations. The notion of system equivalence [27] is under test: what if all observers are not equal and there is a call for a next generation observer theory?

Finally, a serious quantum physicist could ask: "What do we need a theory of the general observer for, when quantum physics works so well and has its tremendous powers in real life?". There is no denying that. A serious perceptual psychologist can answer: "Of course quantum mechanics works, because it is a human science and deals with the world we humans can observe, interpret and manipulate."